\documentclass[11pt,a4paper]{article}
\usepackage{jheppub}
\pdfoutput=1

\usepackage[utf8]{inputenc}
\usepackage{amsmath}
\usepackage{amsfonts}
\usepackage{bm}
\usepackage{tikz}
\usetikzlibrary{arrows}
\usetikzlibrary{shapes}
\usepackage[english]{babel}
\usepackage[autostyle]{csquotes}
\usepackage{braket}
\usepackage{graphics}
\usepackage{tkz-euclide}
\usepackage{tikz-cd}
\usepackage[toc,page]{appendix}
\usetikzlibrary{decorations.markings,arrows}
\usetikzlibrary{arrows,shapes,backgrounds}
\usetikzlibrary{decorations.pathreplacing,decorations.markings}
\usetikzlibrary{calc, trees, positioning, arrows, shapes, shapes.multipart, shadows, matrix, decorations.pathreplacing, decorations.pathmorphing}
\usetikzlibrary{shapes.misc}
\usepackage{hyperref}
\usepackage{bookmark}
\usepackage{verbatim}

\newcommand{\ba}{\begin{equation}\begin{aligned}}
\newcommand{\ea}{\end{aligned}\end{equation}}
\newcommand{\eref}[1]{(\ref{#1})}
\def\be{\begin{equation}}
\def\ee{\end{equation}}

\def\3d{$3d~\mathcal N=4$}

\def\node#1#2{\overset{#1}{\underset{#2}{\circ}}}
\def\noder#1#2{\overset{#1}{\underset{#2}{\textcolor{red}{\circ}}}}
\def\nodeo#1#2{\overset{#1}{\underset{#2}{\textcolor{orange}{\circ}}}}

\def\snode#1#2{\overset{#1}{\underset{#2}{\scriptstyle\square}}}
\def\flav#1{\overset{\scriptstyle#1}{\overset{\square}{\scriptstyle\vert}}}
\def\snode#1#2{\overset{#1}{\underset{#2}{\scriptstyle\square}}}
\def\topnode#1#2{\overset{#1}{\overset{{\displaystyle\circ}\rlap{\,\,$\scriptstyle#2$}}{\scriptstyle\vert}}}
\def\nodesq#1#2{\overset{#1}{\underset{#2}{\boxcircle}}} 
\usepackage{stmaryrd}
\def\nodersq#1#2{\overset{#1}{\underset{#2}{\textcolor{red}{\boxcircle}}}} 
\def\nodeosq#1#2{\overset{#1}{\underset{#2}{\textcolor{orange}{\boxcircle}}}} 
\usepackage{ntheorem}
\newtheorem{theorem}{Claim}

\theoremheaderfont{\normalfont\itshape}

\theoremsymbol{\ensuremath{\blacksquare}} % \theoremsymbol{\ensuremath{_\square}}

\theoremstyle{nonumberplain}
\preprint{Imperial/TP/20/AH/01}
\title{Ungauging Schemes and Coulomb Branches of Non-simply Laced Quiver Theories}
\author[]{Amihay Hanany}
\author[]{and Anton Zajac}
\affiliation[]{Theoretical Physics, The Blackett Laboratory, Imperial College London\\ 
Prince Consort Road, London, SW7 2AZ United Kingdom}
\emailAdd{a.hanany@imperial.ac.uk}
\emailAdd{anton.zajac@imperial.ac.uk}

\abstract{Three dimensional Coulomb branches have a prominent role in the study of moduli spaces of supersymmetric gauge theories with $8$ supercharges in $3,4,5$, and $6$ dimensions. Inspired by simply laced $3$d $\mathcal{N}=4$ supersymmetric quiver gauge theories, we consider Coulomb branches constructed from non-simply laced quivers with edge multiplicity $k$ and no flavor nodes. In a computation of the Coulomb branch as the space of dressed monopole operators, a center-of-mass $U(1)$ symmetry needs to be ungauged. Typically, for a simply laced theory, all choices of the ungauged $U(1)$ (i.e. all choices of \emph{ungauging schemes}) are equivalent and the Coulomb branch is unique. In this note, we study various ungauging schemes and their effect on the resulting Coulomb branch variety. It is shown that, for a non-simply laced quiver, inequivalent ungauging schemes exist which correspond to inequivalent Coulomb branch varieties. Ungauging on any of the long nodes of a non-simply laced quiver yields the same Coulomb branch $\mathcal{C}$. For choices of ungauging the $U(1)$ on a short node of rank higher than $1$, the GNO dual magnetic lattice deforms such that it no longer corresponds to a Lie group, and therefore, the monopole formula yields a non-valid Coulomb branch. However, if the ungauging is performed on a short node of rank $1$, the one-dimensional magnetic lattice is rescaled conformally along its single direction and the corresponding Coulomb branch is an orbifold of the form $\mathcal{C}/\mathbb{Z}_k$. Ungauging schemes of $3$d Coulomb branches provide a particularly interesting and intuitive description of a subset of actions on the nilpotent orbits studied by Kostant and Brylinski \cite{KB92}. The ungauging scheme analysis is carried out for minimally unbalanced $C_n$, affine $F_4$, affine $G_2$, and twisted affine $D_4^{(3)}$ quivers, respectively. The analysis is complemented with computations of the Highest Weight Generating functions.}


\begin{document}
\maketitle

%%%%%%%%%%%%%%%%%%%%%%%%%%%%%%%%%%%%%%%%%%%%%%
\section{Introduction}

Past years have seen a tremendous amount of work directed towards deeper understanding and computations of Coulomb branches of \3d quiver gauge theories \cite{CHZ13,CFHM14,BDG15,N15,BFN16}. Exploitation of $3d$ mirror symmetry \cite{IS96} which allows to use the known description of Higgs branch to understand the Coulomb branch and vice versa, relies on one's ability to compute Coulomb branches of various families of quiver gauge theories (i.e. $T_{\rho}^{\sigma}$ theories, Sicilian theories, minimally unbalanced theories, multiplicity-free varieties, etc.) \cite{Cremonesi:2014uva,BTX10,HPsic17,MUQ18}. Furthermore, $3d$ Coulomb branches have also proven extremely useful in the recent studies of $4d$ \cite{DelZotto:2014kka} $5d$ \cite{FHMZ17,Cremonesi:2015lsa} and $6d$ \cite{Hanany:2018vph,DGHZ18,Hanany:2018cgo,Hanany:2018uhm} Higgs branches. They also play a central role in quiver subtraction \cite{Cabrera:2018ann}, magnetic quivers \cite{Cabrera:2019izd,Cabrera:2019dob}, brane webs \cite{Cabrera:2018jxt,Bourget:2019rtl} and partial Higgs mechanism \cite{Hasse}.\footnote{Note that in many of the mentioned applications, \3d Coulomb branches are used as an abstract construction of a geometric space and need not necessary correspond to a vacuum moduli space of a concrete physical theory.} 

There is a missing technical gap between the results brought to light over the years and a detailed computation of Coulomb branch variety for a given non-simply laced quiver. The gap is related to the choice of \emph{ungauging scheme}. This note addresses the gap and explores the  subtleties  related to the number and structure of admissible Coulomb branches for a non-simply laced quiver theory. The main aim of this note is to establish and provide evidence for the following claim:
\begin{theorem}{\textbf{(Number of Coulomb branches):}} \label{Main Claim}
Every  \3d theory prescribed by a non-simply laced quiver $\sf Q$ with single non-simply laced edge of multiplicity $k$, consisting solely of gauge nodes, admits at least $k_s+1$ different Coulomb branches, where $k_s$ is the number of rank $1$ short nodes of $\sf Q$ modulo outer automorphisms of $\sf Q$.\footnote{Ungauging schemes that gauge fix the residual $U(1)$ on a linear combination of nodes are not considered.}
\end{theorem}

As previously mentioned, we consider quivers with solely gauge nodes and no flavor nodes.\footnote{A natural way to prevent the ungauging scheme ambiguity is to consider quivers with at least one flavor node. In mathematics literature quivers of this type are referred to as \textit{framed quivers} \cite{nakajima1994}. The framing circumvents the ambiguity by declaring which node is ungauged.} The mathematical apparatus used for the computation of the Coulomb branch is built around Hilbert series (HS) \cite{STANLEYhs}, which is a graded generating function which counts all gauge invariant operators in the chiral ring of a gauge theory \cite{Benvenuti:2006qr}. In its unrefined form, it encodes information about the Coulomb branch algebraic variety such as its dimension, degree of generators, number and degree of relations. The unrefined form of the Hilbert series also suffices for an immediate comparison of volumes of two Coulomb branches, providing a necessary test for whether two Coulomb branches are related by an orbifold action. Furthermore, in its refined form (or related forms such as the Highest Weight Generating function (HWG) \cite{Hanany:2014dia} or the modified Hall-Littlewood polynomials (mHL) \cite{mHL14}) Hilbert Series makes the global symmetry (i.e. the isometry) of the Coulomb branch manifest. In particular, the refined form succinctly encodes all the representation content of the chiral ring under the global symmetry. Techniques for counting gauge invariant operators in a chiral ring involve the \emph{monopole formula} \cite{CHZ13} with further utilization of plethystic functions such as the Plethystic exponential (PE), and the Plethystic logarithm  (PL) \cite{FHH07}. In a computation of a Coulomb branch for a flavorless quiver with purely gauge groups, one needs to \emph{ungauge} (decouple) a residual center-of-mass $U(1)$ symmetry. In case of a simply laced quiver, it can be easily shown that the choice of where to ungauge this $U(1)$ is arbitrary and the computed Coulomb branch is invariant with respect to this choice (see Appendix \ref{A}). By the sequel, the following claim holds: 

\begin{theorem}{\textbf{(Number of Coulomb branches of a Simply Laced Quiver):}} \label{Claim 2}
A simply laced (unitary) quiver admits a single unique Coulomb branch.
\end{theorem}

 Since most of existing literature is devoted to the study of simply laced unitary quivers, the `non-uniqueness' of the Coulomb branch, which is manifested by choosing different node of a quiver to be ungauged, has largely been ignored. Despite not playing a crucial role for simply laced theories, in the case of non-simply laced quivers one must pay special attention to what we term as the choice of \emph{ungauging scheme}. Results obtained in this note further indicate that the following claim holds: \\

\begin{theorem}{\textbf{(Orbifold Coulomb Branch for Rank $1$ Short Ungauging Schemes):}} \label{Orbifold Claim}
Let $\sf Q$ be a \3d quiver consisting solely of gauge nodes with a single non-simply laced edge of multiplicity $k$. Further, let $\sf Q$ have $k_s$ rank $1$ short nodes. Let us denote by  $\mathcal{C}_L$ the Coulomb branch corresponding to the ungauging schemes on the long side of the quiver. Then, for every choice of ungauging scheme on the short side of the quiver, such that the ungauged node is of rank $1$, the Coulomb branch has the form
\be
\mathcal{C}_{S} = \mathcal{C}_L/ \mathbb{Z}_k \;, %not S_k ?
\ee
where the precise action of $\mathbb{Z}_k$  on $\mathcal{C}_L$ depends on the choice of the ungauged node.
\end{theorem}
Claim \ref{Orbifold Claim} initiated a broader study of actions on quivers in attempt to understand and reproduce the Kostant-Brylinski results in \cite{KB92} using three-dimensional Coulomb branches. In the present note, ungauging scheme analysis reproduces four of the nine results in \cite{KB92}. These are collected in table \ref{tab:KB} in anticipation, however, the computations and detailed analysis are contained in later sections. It should be noted, that one more result of \cite{KB92} is encountered (see table \ref{tab:KBninth}) in the present note, however, a detailed analysis is contained in the upcoming work \cite{BHMfolding20}. Therein, the authors use the notions of quiver folding \cite{Nakajima:2019olw} and discrete ungauging to understand the remaining results of \cite{KB92}. 

\begin{table}
	\centering
\begin{tabular}{| m{0.43cm} |m{0.65cm}| m{3.4cm} | m{0.43cm} | m{3.2cm}| m{1cm}| m{1cm}| }
		\hline
		KB No. & $\mathfrak{g}'$ & $\sf Q_{G'}$  & $\mathfrak{g}$ & $\sf Q_{G}$ & Action & Dim \\ \hline
			 4 & $E_6$ &
		 $ \node{}1 - \node{}2 - \node{\topnode{\topnode{}1}2}3 - \node{}2 - \node{}1 $ 
		 & $F_4$ &
		  $ \node{}1 - \node{}2 - \node{}3 <= \node{}4 - \node{}2 $
		   & \footnotesize{DU}   &  \footnotesize{$11$} \\ \hline
	 	 5 & $G_2$ &
		  $  \node{\flav 1}2 \equiv > \node{}1 $ 
		 & $A_2$ &
		 $  \node{}1 - \node{}2 \equiv> \snode{}1 $ 
		  &  $\mathbb{Z}_3$ &  \footnotesize{$3$}
	 \\ \hline
	 	 6 & $B_n$ &
		 $ \underbrace{\node{}1 - \node{\flav 1}2 - \cdots - \node{}2 => \node{}1}_{n} $ 
		 & $D_n$ &
		 $ \underbrace{\node{}1 - \node{\topnode{} 1}2 - \cdots - \node{\flav 2}2}_{n-1} $
		  &  $\mathbb{Z}_2$ &  \footnotesize{$2n-2$}
	 \\ \hline
	 	 7 & $F_4$ &
		 $ \node{\flav 1}2 - \node{}3 => \node{}2 -\node{}1 $ 
		 & $B_4$ &
		  $ \node{}1 - \node{}2 - \node{}3 => \node{\flav 1}2  $ 
		  &  $\mathbb{Z}_2$ &  \footnotesize{$8$}
	 \\ \hline	
	\end{tabular}
	\caption{Results $4$,$5$,$6$, and $7$ of Table 1 in \cite{KB92}, corresponding Coulomb branch quivers, actions and dimensions. Round and square nodes denote gauge and flavor groups, respectively. DU denotes an action on quivers termed \emph{discrete ungauging} which is studied in \cite{BHMfolding20}
	}
	\label{tab:KB}
\end{table}  
In \cite{CFHM14} flavored non-simply laced quivers corresponding to single instantons on $\mathbb{C}^2$ \cite{BHN10} are considered. The Coulomb branches obtained in \cite{CFHM14} correspond to ungauging schemes on the affine node of the corresponding affine Dynkin diagram. In \cite{Dey:2016qqp} framed non-simply laced quivers of $B$-type are studied using both the Hilbert Series methods as well as the supersymmetric partition function on a $3$-sphere $S^3$.  \\
 
We demonstrate Claim \ref{Main Claim} and Claim \ref{Orbifold Claim} using the example of a minimally unbalanced $C_3$ quiver in section \ref{2}. The same program is applied to the affine $B_3$ and affine $F_4$ quivers in section \ref{3} and \ref{4}, respectively.  As a natural next step, in sections \ref{5} and \ref{6} the analysis is extended to quivers with a triple-laced edge using the affine $G_2$ and twisted affine $D_4^{(3)}$ quivers, respectively. In section \ref{7} the patterns found in section \ref{2} are used to derive the form of the Coulomb branch HWG for an infinite sequence of minimally unbalanced $C_n$ quivers. % get rid of orbi-lienti-folding?
%Section \ref{6} contains a discussion of the two-way relations between quivers related by orbifolding and orientifolding. 
Section \ref{conclusions} contains a summary of the results as well as a discussion of possible implications of Claim \ref{Main Claim} and Claim \ref{Orbifold Claim} in the subsequent study of quiver theories. Appendix \ref{A} contains a brief review of the formulae used in the computation of Coulomb branches. The relationship between the choice of ungauging scheme and the conformal dimension, hence, the resulting Coulomb branch, is discussed in Appendix \ref{B}.

%%%%%%%%%% C3
\section{Ungauging Schemes for $C_3$} \label{2}

Lets begin with the minimally unbalanced $C_3$ quiver depicted in equation \ref{eq:c3-1} which has an extra unbalanced node (drawn red) connected to the rank $3$ long node such that its excess is $e=-1$.\footnote{Excess of a node is defined as $e=N_f - 2N_c$, where $N_f$ is the number of flavors and $N_c$ is the number of colors. Quivers with all nodes balanced except for one node with non-zero excess are termed minimally unbalanced. For a complete classification of minimally unbalanced quiver gauge theories refer to \cite{MUQ18}. See also \eref{eq:balance} in Appendix \ref{A}.} 

\begin{equation} \label{eq:c3-1}
\node{}1-\node{}2 <= \node{}3-\noder{}2
\end{equation}

The two nodes on the left in equation \ref{eq:c3-1} are \emph{short} and the two nodes on the right are \emph{long}, respectively. Since all nodes of the quiver are gauge nodes, in the computation of the Coulomb branch, one of the magnetic charges is set to zero. This is known as the \emph{ungauging} or \emph{decoupling} of the center-of-mass $U(1)$. The most natural choice is to ungauge on the unbalanced red node since the remaining balanced part of the quiver forms the $C_3$ Dynkin diagram which corresponds to the expected global symmetry on the Coulomb branch \cite{DGHZ18}. The ungauged node shall always be denoted by a \emph{squircle} $\nodesq{}{} $.\footnote{When one ungauges on a rank $1$ node it follows that the whole node is ungauged, hence, it becomes a flavor node denoted as $\snode{}{}$. On the other hand, ungauging on a node with rank $r>1$ fixes the origin in the space of magnetic charges (i.e. introduces a delta function on one of the components of the magnetic flux at the corresponding node). Contrary to the more common use of the word squircle, referring to a square with rounded corners, hereby authors mean a symbiotic co-existence of a circle and a square at a given node position.}

 After declaring which node is ungauged (i.e. which node becomes a squircle) one says that a particular \emph{ungauging scheme} is chosen. Let us begin by ungauging on the rightmost long node. For such choice of the ungauging scheme the quiver is given in equation \ref{eq:c3-2}.

\begin{equation} \label{eq:c3-2}
\node{}1-\node{}2 <= \node{}3-\nodersq{}2
\end{equation}

Computation of the unrefined Hilbert series and the corresponding Coulomb branch for the quiver in \ref{eq:c3-2} yields
\be \label{eq:1-1}
HS(t)=\frac{1}{(1-t)^{14}}, \quad \mathcal{C}=\mathbb{H}^7.
\ee
One sees that the Coulomb branch is a freely generated algebraic variety of quaternionic dimension $7$. Typically, when the excess is $e=-1$ and the unbalanced node is connected to a Dynkin node corresponding to a pseudo-real representation, the Coulomb branch is freely generated and one finds a certain embedding. In the case at hand one has:
\be
[0,0,1]_{Sp(3)} \hookleftarrow [1,0,0,0,0,0,0]_{Sp(7)},
\ee
(i.e. the mapping of the $14$-dimensional pseudo-real fundamental rep of $Sp(7)$ into the $14$-dimensional $3$-rd rank antisymmetric pseudo-real representation of $Sp(3)$, corresponding to the Dynkin node to which the unbalanced node is attached). For the quiver in equation \ref{eq:c3-2} the Highest Weight Generating function (HWG) can be written in terms of the $Sp(7)$ highest weight fugacities $[\mu_1,\dots , \mu_7]$ and using the Plethystic Exponential as
\be 
HWG=PE\left[\mu_1t \right],
\ee
or alternatively, as
\be \label{eq:hwgC3}
HWG=PE\left[\mu_1^2 t^2  +\mu_2^2 t^4 +t^4 + \mu_3 t +\mu_3 t^3 \right],
\ee
where $[\mu_1,\mu_2,\mu_3]$ denote the highest weight fugacities for $Sp(3)$. The Highest Weight Generating function \ref{eq:hwgC3} is revisited in the derivation of the general case in section \ref{6}. It can be also obtained from equation (23) in \cite{HPsic17} by thinking of \ref{eq:c3-2} as the folded version of the quiver in Figure 4 therein for $N=3$. By inspection of \ref{eq:hwgC3} at order $t$ one recognises the $3$-rd rank antisymmetric rep of $Sp(3)$ corresponding to the node where the unbalanced node with $e=-1$ attaches. The balanced part of the quiver contributes with the adjoint rep of $Sp(3)$ at order $t^2$, making the $C_3$ global symmetry manifest. Working out the $2$-nd, $3$-rd and $4$-th symmetric product of $\mu_3$:
\begin{align}
Sym^2 \mu_3 = \mu_1^2 + \mu_3^2 \quad\quad\quad\quad\quad\quad\quad\quad\quad \;  \\
Sym^3 \mu_3 = \mu_3^3 + \mu_1^2 \mu_3 +\mu_3 \quad\quad\quad\quad\quad\quad \\
Sym^4 \mu_3 = \mu_3^4 + \mu_1^2 \mu_3^2 +\mu_1^4 +\mu_3^2 +\mu_2^2 +1
\end{align}
reveals the presence of the singlet at order $t^4$, and hence, justifies expression \ref{eq:hwgC3}. Let us now demonstrate the effect of choosing a different ungauging scheme for the quiver in equation \ref{eq:c3-1}. For this purpose, compute the Coulomb branch for the quiver depicted in equation \ref{eq:c3-3}, where we choose to ungauge on the rank $3$ long node. Again, the ungauged node is denoted by a squircle.
\begin{equation} \label{eq:c3-3}
\node{}1-\node{}2 <= \nodesq{}3-\noder{}2
\end{equation}
Claim \ref{Main Claim} implies that the Coulomb branch should be the same as in the previous case since the new ungauging scheme remains on the long side of the quiver. Indeed, the computation yields:
\be \label{eq:1-2}
HS(t)=\frac{1}{(1-t)^{14}}, \quad \mathcal{C}=\mathbb{H}^7.
\ee
Equality of \ref{eq:1-2} and \ref{eq:1-1} is in accord with Claim \ref{Main Claim}! So far, we have found two identical Coulomb branches for the two long ungauging schemes. Let us now consider a scenario depicted in equation \ref{eq:c3-4}, where the leftmost short node is ungauged, and accordingly, denoted by a squircle.
\begin{equation} \label{eq:c3-4}
\nodesq{}1-\node{}2 <= \node{}3-\noder{}2
\end{equation}
In this case one computes the unrefined Hilbert series to be
\be \label{eq:1-3}
HS(t)= \frac{1 + 6 t^2 + t^4}{(1 - t)^{10} (1 - t^2)^4} ,
\ee
which clearly describes a different Coulomb branch! From the first term in the denominator, observe that the computed Coulomb branch variety has a $5$-dimensional free part. Furthermore, there is a non-trivial part corresponding to a $\mathbb{C}^4/\mathbb{Z}_2$ singularity such that the $\mathbb{Z}_2$ action naturally acts on all the coordinates of $\mathbb{C}^4$. To show the action explicitly, start with the HWG for the freely generated $\mathbb{C}^4$
\be
HWG_{\; \mathbb{C}^4}=PE\left[\mu_1t\right] ,
\ee 
where $\mu_1$ is the highest weight fugacity of $Sp(2)$. Next, construct the $\mathbb{Z}_2$ projection 
\be \label{eq:sp1instanton}
 HWG_{\: \mathbb{C}^4/\mathbb{Z}_2} = \frac{1}{2} \left(PE\left[\mu_1t\right]+PE\left[-\mu_1t\right] \right)
=PE\left[\mu_1^2 t^2\right] \;,
\ee
resulting in the HWG corresponding to $\mathbb{C}^4/\mathbb{Z}_2$ singularity. Indeed, \ref{eq:sp1instanton} describes a moduli space of one $Sp(2)$ instanton on $\mathbb{C}^2$ \cite{BHN10}. The Coulomb branch obtained for the ungauging scheme depicted in \ref{eq:c3-4} takes the form:
\be
\mathcal{C}=\mathbb{H}^5 \times \mathbb{C}^4 / \mathbb{Z}_2 .
\ee

 Finally, consider the last ungauging scheme by letting the rank $2$ short node be ungauged. The corresponding ungauging scheme is shown in equation \ref{eq:c3-5}. Recall, that the ungauging scheme fixes the origin of the two-dimensional magnetic lattice at the associated node. This is achieved by introducing a delta function which sets to zero one of the magnetic charges at the node. 
% Start with an interacting theory and flow to the IR. New monopole operators arise that are either free or in form of an orbifold.... nice computation

\begin{equation} \label{eq:c3-5}
\node{}1-\nodesq{}2 <= \node{}3-\noder{}2
\end{equation}

The unrefined HS computed in this case takes the form:
\be \label{eq:1-4}
HS(t)=\frac{(1 + t^2) (1 + 6 t + 18 t^2 + 28 t^3 + 38 t^4 + 28 t^5 + 18 t^6 + 
   6 t^7 + t^8)}{(1 - t)^{14} (1 + t)^6 (1 + t + t^2)^2},
\ee 
and we see that, indeed, it differs from both Hilbert series \ref{eq:1-3} for the quiver in equation \ref{eq:c3-4} as well as from expression \ref{eq:1-1} (resp. expression \ref{eq:1-2}).

 It is important to note that the variety described by Hilbert series \ref{eq:1-4}, although still being a Gorenstein singularity \cite{MR0224620}, does not resemble any known form of a hyperK\"ahler moduli space. In this note, we encounter the same problem every time the choice of ungauging scheme involves ungauging on a short node with rank $r>1$. This problem stems from the fact that for such ungauging schemes, the monopole formula summation runs over a lattice which is non-conformally scaled (i.e. scaled by $2$ in one of its dimensions) and no longer corresponds to the GNO dual lattice of the gauge group at the given node (i.e. $U(2)$ in the present case). See the derivation in Appendix \ref{B}. \\

In summary, we obtained two different Coulomb branches described by the Hilbert series in equations \ref{eq:1-1} (resp. \ref{eq:1-2}), and \ref{eq:1-3}, respectively.  Hilbert series \ref{eq:1-1} (and \ref{eq:1-2}) describes a freely generated Coulomb branch. Hilbert series \ref{eq:1-3} describes a Coulomb branch that is a $\mathbb{Z}_2$ orbifold of the former. The number of Coulomb branches for the quiver in \ref{eq:c3-1} indeed equals
\be
k_s +1 = 2 ,
\ee 
where $k_s$ is the number of short rank $1$ nodes of the quiver. Furthermore, on the long side of the quiver, the position of the ungauged node can be arbitrary and one computes the same Coulomb branch $\mathcal{C}_L$. Table \ref{tab:C3Tab} collects the unrefined expansions of the Plethystic logarithm for the various choices of ungauging schemes. Note that table \ref{tab:C3Tab} contains only one of the equivalent long ungauging schemes. All the results in this section suggest the validity of Claim \ref{Main Claim} and Claim \ref{Orbifold Claim}.
\begin{table}
	\centering
	\begin{tabular}{|c|c|}
		\hline
		Ungauging scheme & Unrefined PL  \\ \hline	
$\node{}1-\node{}2 <= \node{}3-\nodersq{}2$ & $PL=14t$ \\ \hline 		
$\nodesq{}1-\node{}2 <= \node{}3-\noder{}2$ & $PL=10 t + 10 t^2 - 20 t^4 + 64 t^6 - 280 t^8+ o(t^9)$ 
\\ \hline \hline
$\node{}1-\nodesq{}2 <= \node{}3-\noder{}2$ & $PL=12 t + 4 t^2 - 8 t^3 + 31 t^4 - 86 t^5 + 147 t^6 - 32 t^7 - 813 t^8+ o(t^9)$ 
\\ \hline
	\end{tabular}
	\caption{Different choices of ungauging schemes for the minimally unbalanced $C_3$ quiver. The ungauging scheme in the first row has a simple refined PL given by ${PL}_{ref}= [0,0,1]_{Sp(3)}t$ in terms of the Dynkin labels of $Sp(3)$. The ungauging in the last row yields a non-valid Coulomb branch.}
	\label{tab:C3Tab}
\end{table}

% COnstruct PE of 3rd rank ASYm of C3 [0,0,1]t do projection to C2xC1 and compute HWG

%%%%%%%%%%% B3 
\section{Ungauging Schemes for $B_3$} \label{3}

In this section, we consider the affine $B_3$ quiver given by \ref{eq:B31}. 
\be \label{eq:B31}
\node{}1 - \node{\topnode{} 1}2  => \node{}1
\ee
It enjoys a $\mathbb{Z}_2$ outer automorphism symmetry which rotates the rank $1$ `fork' nodes. There is one rank $1$ short node, hence, $k_s=1$ and one is to find two inequivalent ungauging schemes. The results of the Coulomb branch computation for each ungauging scheme are collected in table \ref{tab:B3}.
\begin{table}
	\centering
	\begin{tabular}{|c|c|c|}
		\hline
		Ungauging scheme & Hilbert Series, $HS(t)$ & Coulomb branch, $\mathcal{C}$ \\ \hline
$\nodesq{}1 - \node{\topnode{} 1}2  => \node{}1$ & $\frac{1 + 13 t^2 + 28 t^4 + 13 t^6 + t^8 }{(1 - t^2)^{8}}$ & $\overline{min}_{B_3}$\\ \hline
 $\node{}1 - \nodesq{\topnode{} 1}2  => \node{}1$ & $\frac{1 + 13 t^2 + 28 t^4 + 13 t^6 + t^8 }{(1 - t^2)^{8}}$ & $\overline{min}_{B_3}$\\ \hline
  $\node{}1 - \node{\topnode{} 1}2  => \nodesq{}1$ & $\frac{(1+t^2)^2 (1 + 5 t^2 +t^4)}{(1 - t^2)^{8}}$ & $\overline{n.min}_{D_3} $\\ \hline
	\end{tabular}
	\caption{Different ungauging schemes for the affine $B_3$ quiver.}
	\label{tab:B3}
\end{table}
The Coulomb branches in the first two rows of table \ref{tab:B3} correspond to the closure of the minimal nilpotent orbit of $\mathfrak{so}(7)$ and the highest weight generating function is given by \cite{HK16}:
\be
HWG_{\; \overline{min}_{B_3}}=PE\left[  \mu_2 t^2 \right] ,
\ee
where $\mu_2$ denotes the highest weight fugacity for the adjoint representation of $SO(7)$. The last row of Table \ref{tab:B3} corresponds to the closure of the next-to-minimal nilpotent orbit of $D_3 \cong A_3$ \cite{HK16}. Denote by $HS(t)_L$ and $HS(t)_S$ the Hilbert series in the first two and in the last row of Table \ref{tab:B3}, respectively. Comparison of the volumes of the corresponding varieties yields
\be \label{eq:B3volumes}
\frac{HS(t)_L \mid_{t\rightarrow 1} \sim \frac{R_L}{(1-t)^8}}{HS(t)_S \mid_{t\rightarrow 1} \sim \frac{R_S}{(1-t)^8}} =\frac{\frac{7}{32}}{\frac{7}{64}}=2= \text{ord}(\mathbb{Z}_2)
\ee
where $R_L, R_S$ denote the associated residues at $t=1$ and ord() denotes the order of a group. Expression \ref{eq:B3volumes} indicates\footnote{Comparison of the volumes of two Coulomb branches is a necessary but not sufficient check of a particular orbifold relation between them.}
\be \label{eq:KBB3}
 \overline{n.min}_{D_3} =\overline{min}_{B_3}/\mathbb{Z}_2 .
\ee 
To see the $\mathbb{Z}_2$ action explicitly, first decompose the highest weight fugacity of the adjoint representation of $B_3$ into the highest weight fugacities of $A_3$:
\be
\mu_2 \longrightarrow \mu_1 \mu_3 + \mu_2 .
\ee
Then, the $\mathbb{Z}_2$ projection is constructed as
\be
\frac{1}{2} \left( 
  \frac{1}{\left(1-\mu_1\mu_3 t^2\right) \left(1- \mu_2 t^2\right)} + \frac{1}{\left(1-\mu_1\mu_3 t^2\right)\left(1+\mu_2 t^2\right) } \right) =PE\left[\mu_1\mu_3 t^2+ \mu_2^2 t^4\right],
\ee
where on the right hand side, one recognizes the HWG for $\overline{n.min}_{D_3} \cong \overline{n.min}_{A_3}$. This justifies equation \ref{eq:KBB3}, which is the sixth Kostant-Brylinski \cite{KB92} result (for $n=3$) advertised in table \ref{tab:KB}. In summary, results in this section are in accord with both Claim \ref{Main Claim} as well as Claim \ref{Orbifold Claim}.

%%%%%%%%%% F4
\section{Ungauging Schemes for $F_4$} \label{4}

In this section, we study the quiver given by \ref{eq:f4} which is in the form of the affine Dynkin diagram of $F_4$.
\be \label{eq:f4}
\nodeo{}1-\node{}2-\node{}3 => \node{}2-\node{}1
\ee
The extra affine node (orange)\footnote{Extra balanced nodes are drawn orange in order to be distinguished from extra unbalanced nodes (red).} is connected to the adjoint Dynkin node such that its excess is zero and the whole quiver is balanced. There is a single rank $1$ short node, hence, $k_s=1$. Moreover, the quiver lacks any outer automorphism symmetry, thus, based on Claim \ref{Main Claim}, admits $2$ different Coulomb branches. The various choices of the ungauging schemes as well as the resulting Hilbert series and Coulomb branches are collected in table \ref{tab:F4}. \\
%Some of these results were initially computed as part of \cite{SW}.
\begin{table}
	\centering
	\begin{tabular}{|c|c|c|}
		\hline
		Ungauging scheme & Hilbert Series, $HS(t)$ & Coulomb branch, $\mathcal{C}$ \\ \hline
$\nodeosq{}1-\node{}2-\node{}3 => \node{}2-\node{}1$ & $\frac{1 + 36 t^2 + 341 t^4 + 1208 t^6 + 1820 t^8 + 1208 t^{10} + 341 t^{12} + 
 36 t^{14} + t^{16}}{(1 - t^2)^{16}}$ &  $\overline{min}_{F_4}$\\ \hline
$\nodeo{}1-\nodesq{}2-\node{}3 => \node{}2-\node{}1$ & $\frac{1 + 36 t^2 + 341 t^4 + 1208 t^6 + 1820 t^8 + 1208 t^{10} + 341 t^{12} + 
 36 t^{14} + t^{16}}{(1 - t^2)^{16}}$ & $\overline{min}_{F_4}$\\ \hline
$\nodeo{}1-\node{}2-\nodesq{}3 => \node{}2-\node{}1$ & $\frac{1 + 36 t^2 + 341 t^4 + 1208 t^6 + 1820 t^8 + 1208 t^{10} + 341 t^{12} + 
 36 t^{14} + t^{16}}{(1 - t^2)^{16}}$ &  $\overline{min}_{F_4}$\\ \hline \hline
$\nodeo{}1-\node{}2-\node{}3 => \nodesq{}2-\node{}1$  & $\frac{(1 + t^2)^2 (1 + 26 t^2 + 149 t^4 + 272 t^6 + 149 t^8 + 26 t^{10} + 
   t^{12})}{(1 - t^2)^{16}}$ &  non-valid\\ \hline
$\nodeo{}1-\node{}2-\node{}3 => \node{}2-\nodesq{}1$ & $\frac{1 + 20 t^2 + 165 t^4 + 600 t^6 + 924 t^8 + 600 t^{10} + 165 t^{12} + 
 20 t^{14} + t^{16}}{(1 - t^2)^{16}}$ &  $\overline{n.n.min}_{B_4}$ \\ \hline
	\end{tabular}
	\caption{Different ungauging schemes for the affine $F_4$ quiver.}
	\label{tab:F4}
\end{table}
In accord with Claim \ref{Main Claim} one observes that all three ungauging schemes on the long side of the quiver, grouped in the first three rows of table \ref{tab:F4}, correspond to the same Coulomb branch. In all of these cases the Coulomb branch is the closure of the minimal nilpotent orbit of $\mathfrak{f}_4$ algebra which is known to correspond to the reduced moduli space of one $F_4$ instanton on $\mathbb{C}^2$ \cite{CFHM14,Hanany:2017ooe}. The resulting HS matches (5.44) in \cite{BHN10} and the HWG is known to have the form:
\be
HWG_{\; \overline{min}_{F_4}}=PE\left[\mu_1 t^2\right],
\ee
where $\mu_1$ denotes the adjoint highest weight fugacity of $F_4$. The remaining two cases, grouped in the last two rows of table \ref{tab:F4}, result from ungauging schemes on the short nodes. In particular:
 \begin{itemize}
 \item For the ungauging scheme in the fourth row of table \ref{tab:F4}, we encounter the same problem as in the previous sections due to the non-conformal scaling of the GNO dual lattice and the space computed by monopole formula techniques is not a valid Coulomb branch.
  \item The ungauging scheme in the last row of table \ref{tab:F4} yields a quiver which enjoys $B_4$ global symmetry and the Coulomb branch corresponds to the closure of the $16$-dimensional next-to-next-to minimal nilpotent orbit of $\mathfrak{so}(9)$ algebra \cite{HK16}. To see the orbifold action explicitly, first inspect that the $F_4$ adjoint highest weight fugacity decomposes as
  \be
  \mu_1 \longrightarrow \mu_2 + \mu_4 ,
  \ee
  where on the right hand side, $\mu_2, \mu_4$ are the fugacities for the highest weights of $B_4$. Next, construct the explicit $\mathbb{Z}_2$ projection
  \be \label{eq:derivedb4}
\frac{1}{2} \left( 
\frac{1}{(1-\mu_2 t^2)( 1- \mu_4 t^2)} + \frac{1}{(1-\mu_2 t^2)(1+\mu_4 t^2)} 
\right) 
=PE\left[\mu_2 t^2+ \mu_4^2 t^4\right].
\ee
Right hand side of expression \ref{eq:derivedb4} is the HWG for the $\overline{n.n.min}_{B_4} $ orbit given in terms of $B_4$ highest weight fugacities $[\mu_1,\mu_2,\mu_3,\mu_4]$.
  \end{itemize}
 The results in this section are in accord with Claim \ref{Main Claim} and in further support of Claim \ref{Orbifold Claim} which is related to the orbifold structure of the Coulomb branch for short rank $1$ ungauging schemes. Both the comparison of the volumes of the two Coulomb branches corresponding to the first three versus the last row of table \ref{tab:F4} analogous to \ref{eq:B3volumes} as well as the explicit projection in \ref{eq:derivedb4} justify the orbifold relation:
 \be \label{eq:KB7}
 \overline{n.n.min}_{B_4} = \overline{min}_{F_4}/ \mathbb{Z}_2 .
 \ee 
Equation \ref{eq:KB7} is the seventh result of Kostant and Brylinski \cite{KB92}. Some unrefined results in this section appear in \cite{SW,MH}.

%%%%%%%%%%% G2
\section{Ungauging Schemes for $G_2$} \label{5}

As a natural step, we now extend our analysis to quivers with a triple laced edge. Let us first study the affine $G_2$ quiver given by \ref{eq:g2quiver}. Since $k_s=1$, and with the lack of outer automorphism symmetry, two different Coulomb branches are expected to exist.
\be \label{eq:g2quiver}
\nodeo{}1-\node{}2 \equiv > \node{}1
\ee
 The summary of results for various ungauging schemes is given in table \ref{tab:G2min}. Let us analyze the three different ungauging schemes contained in table \ref{tab:G2min}:

 \begin{table}
	\centering
	\begin{tabular}{|c|c|c|}
		\hline 
		Quiver & Hilbert Series, $HS(t)$ & Coulomb branch , $\mathcal{C}$\\ \hline  
$\nodeosq{}1-\node{}2 \equiv > \node{}1 $ &$\frac{(1 + t^2) (1 + 7 t^2 + t^4)}{(1 - t^2)^6}$ & $\overline{min}_{G_2}$\\ \hline
$\nodeo{}1-\nodesq{}2 \equiv > \node{}1 $ & $ \frac{(1 + t^2) (1 + 7 t^2 + t^4)}{(1 - t^2)^6}$ & $\overline{min}_{G_2}$\\ \hline		
$\nodeo{}1-\node{}2 \equiv > \nodesq{}1 $ & $ \frac{(1 + t^2) (1 +  t^2+t^4)}{(1 - t^2)^6}$ & $\overline{min}_{G_2}/\mathbb{Z}_3 = \overline{max}_{A_2}$ \\ \hline 
	\end{tabular}
	\caption{Ungauging schemes for the affine $G_2$ quiver.}
		\label{tab:G2min}
\end{table} 
\begin{itemize}
\item For the ungauging on one of the long nodes in the first row of table \ref{tab:G2min}, the Coulomb branch corresponds to the closure of the minimal nilpotent orbit of $\mathfrak{g}_2$ algebra, hence, to the reduced moduli space of one $G_2$ instanton on $\mathbb{C}^2$ \cite{BHN10,CFHM14,Hanany:2017ooe}. The highest weight generating function has the form:
\be \label{eq:hwgg2instanton}
HWG_{\; \overline{min}_{G_2}}=PE \left[  \mu_1t^2 \right] ,
\ee
where $\mu_1$ is the fugacity for the highest weight of the adjoint representation of $G_2$.
\item Ungauging scheme in the second row yields the same Coulomb branch described by the same Hilbert Series. This is in accordance with the prediction of Claim \ref{Main Claim}.
\item Finally, third row depicts the ungauging scheme in which the short node is ungauged. According to Claim \ref{Orbifold Claim} the Coulomb branch takes the form of $\mathbb{Z}_3$ orbifold. To see this explicitly, start with HWG \ref{eq:hwgg2instanton} and decompose the highest weight of the adjoint representation of $G_2$:
\be
\mu_1 \longrightarrow  \mu_1\mu_2  + \mu_1  + \mu_2 ,
\ee
where on the right side, $\mu_1,\mu_2$ are the highest weights fugacities of $A_2$. The $\mathbb{Z}_3$ projection is constructed as:
\be \label{eq:z3average}
 \begin{split}
\frac{1}{3} & \{ PE\left[  \left(\mu_1\mu_2  + \mu_1  + \mu_2 \right)t^2\right]  + \\
&  PE\left[  \left(\mu_1\mu_2 + \omega \mu_1  +  \omega^{-1}  \mu_2\right)t^2\right] + \\
&  PE\left[ \left(\mu_1\mu_2  + \omega^{-1}\mu_1  + \omega \mu_2\right)t^2\right] \},
 \end{split}
\ee
% where ^3=1, identity: 
where $\omega$ employs the cyclic $\mathbb{Z}_3$ action, satisfying $\omega^3 =1$. After the $\mathbb{Z}_3$ projection the Highest Weight Generating function for the last row of table \ref{tab:G2min} is obtained:
 \be
 HWG_{\; \overline{max}_{A_2}} =PE\left[\mu_1 \mu_2 t^2 +\mu_1 \mu_2 t^4 +\mu_1^3 t^6 \\
  +\mu_2^3 t^6  -\mu_1^3 \mu_2^3 t^{12}  \right] ,
 \ee
 where $\mu_1, \mu_2$ denote the fugacities for the highest weights of $A_2$.
 % ASSK! In fact, construction \ref{eq:z3average} amounts to the gauging of the center of $\mathfrak{a}_2$ (or g2?) algebra. 
\end{itemize}
One can also work out the comparison of the volumes of the two valid Hilbert series in table \ref{tab:G2min} in a similar fashion as in \cite{DGHZ18} and \ref{eq:B3volumes} in section \ref{3}: 
\be \label{eq:polesG2}
\frac{\text{vol}({HS}_S)}{\text{vol}({HS}_{L})}=\frac{R_S}{R_L} = \frac{1}{3} = \frac{1}{\text{ord}(\mathbb{Z}_3)},
\ee
where $\text{ord}()$ denotes the order of the group. Subscripts $S$ and $L$ in equation \ref{eq:polesG2} denote Hilbert Series corresponding to the short versus long ungauging schemes, respectively. The analysis shows that $\mathcal{C}_S$ is isomorphic to the closure of the maximal nilpotent orbit of $\mathfrak{sl}_3$ algebra, denoted by $\overline{max}_{A_2}$:
\be \label{eq:KB92resultg2}
 \overline{max}_{A_2} =\overline{min}_{G_2}/\mathbb{Z}_3 .
\ee
Equation \ref{eq:KB92resultg2} reproduces the third result of Kostant and Brylinski advertised in table \ref{tab:KB} in the introduction. This nicely demonstrates that the quiver in the third row of table \ref{tab:G2min} is equivalent to the self mirror dual quiver for the closure of the maximal nilpotent orbit of $A_2$:
\be
\node{}1 - \node{}2 - \snode{}3
\ee
Results \ref{eq:polesG2} and \ref{eq:KB92resultg2} provide further evidence for Claim \ref{Orbifold Claim}. In this section, we obtained two different Coulomb branches in accord with Claim \ref{Main Claim}. Some unrefined results in this section appear in \cite{SW,MH}. \\

%%%%%%%%%%% twisted affine D4
\section{Ungauging Schemes for $D_4^{(3)}$} \label{6}

Now, consider the twisted affine $D_4^{(3)}$ quiver depicted in \ref{eq:tad4}. The quiver lacks any outer automorphism symmetry and $k_s=1$ therefore two valid Coulomb branches are anticipated.
\be \label{eq:tad4}
\nodeo{}1 - \node{}2 < \equiv  \node{}3
\ee
The three different ungauging schemes and the corresponding HS and $\mathcal{C}$ are collected in table \ref{tab:D4twisted}.
\begin{table}
	\centering
	\begin{tabular}{|c|c|c|}
		\hline 
		Quiver & Hilbert Series, $HS(t)$ & Coulomb branch, $\mathcal{C}$ \\ \hline  
		$\nodesq{}3 \equiv > \node{}2 - \nodeo{}1 $ &$ \frac{(1 + t^2) (1 + 17 t^2 + 48 t^4 + 17 t^6 + t^8)}{(1 - t^2)^{10}} $ & $\overline{min}_{D_4}$\\ \hline
		$\node{}3 \equiv > \nodesq{}2 - \nodeo{}1 $ &$ \frac{1 + 12 t^2 + 25 t^4 + 36 t^6 + 25 t^8 + 
 12 t^{10} + t^{12}}{(1 - t^2)^{10} (1 + t^2)}$ & non-valid\\  \hline
$\node{}3 \equiv > \node{}2 - \nodeosq{}1 $ &$ \frac{(1 + t^2) (1 + 3 t+ 6 t^2 + 3 t^3 + t^4)(1 - 3 t+ 6 t^2 - 3 t^3 + t^4) }{(1 - t^2)^{10}}$ & $\mathcal{C}_{D_4^{(3)}}= \overline{min}_{D_4}/\mathbb{Z}_3$ \\ \hline
	\end{tabular}
	\caption{Ungauging schemes for the twisted affine $D_4^{(3)}$ quiver.}
		\label{tab:D4twisted}
\end{table}
 In particular, the analysis shows:
%Calculate D4, take all stelar fungs to be the same and compare with [1]-2<triple 3. Find the quotients.
 %1-2<triple 3 refined HS, take all 
%% paper with 1-2-3adj? cite!
\begin{itemize}
\item The ungauging scheme on the long side of the quiver, depicted in the first row of table \ref{tab:D4twisted}, yields a Coulomb branch denoted by $\overline{min}_{D_4}$ which is known to correspond to the closure of the minimal nilpotent orbit of $\mathfrak{so}(8)$ algebra \cite{BHN10}. Moreover, in terms of the Highest Weight Generating function we have
\be \label{eq:hwgd4min}
HWG_{\; \overline{min}_{D_4}}=PE\left[\mu_2t^2\right] ,
\ee
where $\mu_2$ denotes the highest weight fugacity for the adjoint of $D_4$.

\item Ungauging scheme in the second row of table \ref{tab:D4twisted}, with ungauging on the rank $2$ short node, produces a non-valid Coulomb branch as one again encounters the problem with the non-conformal deformation of the GNO dual lattice of the $U(2)$ gauge group.
\item Ungauging scheme depicted in the third row of table \ref{tab:D4twisted} produces a five-dimensional Coulomb branch denoted by $\mathcal{C}_{D_4^{(3)}}$. This is a variety of characteristic height $3$, therefore it is beyond the characteristic height $2$ family of nilpotent orbits \cite{Hanany:2017ooe}. The obtained space is included in the Achar-Henderson analysis (see Table 6 in \cite{achar2011geometric}). In order to derive the HWG, first decompose the highest weight fugacity of the adjoint representation of $D_4$:
\be
\mu_2 \longrightarrow \nu_1+ \nu_2+ \nu_2
\ee
where on the right hand side $\nu_1, \nu_2$ denote the highest weight fugacities of $G_2$.  After expressing the HWG \ref{eq:hwgd4min} in terms of the $G_2$ fugacities, there is also an adjoint contribution appearing at order $t^4$:
%\footnote{This can be peaked into by a simple computation of the dimension of the Coulomb branch from the Highest Weight Generating function.}
\be
PE\left[\mu_2 t^2 \right] \longrightarrow PE\left[ \nu_1 t^2 + \nu_2 t^2 + \nu_2 t^2 + \nu_1 t^4 \right] .
\ee
Next, construct the $\mathbb{Z}_3$ projection explicitly
\be \label{eq:d4twistedaverage}
\begin{split}
\frac{1}{3} & \{ PE \left[ \left(\nu_1+ \nu_2 + \nu_2 \right) t^2 + \nu_1 t^4 \right]  + \\
& PE \left[ \left(\nu_1+ \omega \nu_2+ \omega^{-1} \nu_2 \right) t^2+ \nu_1 t^4 \right] +  \\
& PE \left[ \left(\nu_1+ \omega^{-1} \nu_2 + \omega \nu_2\right) t^2+ \nu_1 t^4 \right] \} ,
\end{split}
\ee
where $\omega$, satisfying $\omega^3 =1$, employs the $\mathbb{Z}_3$ action similarly to \ref{eq:z3average}. Evaluating \ref{eq:d4twistedaverage} yields the Highest Weight Generating function for $\mathcal{C}_{D_4^{(3)}}$ in terms of the fugacities for the highest weights of $G_2$:
\be \label{eq:tad4hwg}
HWG_{\mathcal{C}_{D_4^{(3)}}}=PE[\nu_1t^2 + \nu_1 t^4 +\nu_2^2 t^4  + 2\nu_2^3 t^6 - \nu_2^6  t^{12}] .
\ee

Since the space $\mathcal{C}_{D_4^{(3)}}$ appears for the first time in the context of chiral rings, we include the expansion of the refined PL for completeness. It is given in terms of $G_2$ Dynkin labels $[\nu_1,\nu_2]$ in expression \ref{eq:refpld43} .
\be \label{eq:refpld43}
\begin{split}
PL =&  [1,0]t^2 + \left([1,0]-[0,0] \right)t^4 - \left( [1,0]+[0,1]+[0,2]+[0,0] \right)t^6- \\
 & \left( [1,0]+[0,1] -[2,0] \right)t^8 + O(t^9)
\end{split}
\ee

Both the necessary volume comparison, similar to \ref{eq:B3volumes} and \ref{eq:polesG2}, as well as the explicit HWG computation \ref{eq:d4twistedaverage} imply the orbifold relation \ref{eq:ENO}.
\be \label{eq:ENO}
\mathcal{C}_{D_4^{(3)}}= \overline{min}_{D_4}/\mathbb{Z}_3
 \ee
Equation \ref{eq:ENO} relates the obtained Coulomb branch to the closure of the minimal nilpotent orbit of $\mathfrak{so(8)}$ (first row of table \ref{tab:D4twisted}). This closely resembles the ninth results of \cite{KB92} shown in table \ref{tab:KBninth}, however, the action in our case is $\mathbb{Z}_3$ instead of $S_3$.
\begin{table} \label{tab:KBninth}
	\centering
\begin{tabular}{| m{0.43cm} |m{0.65cm}| m{3cm} | m{0.43cm} | m{3cm}| m{0.5cm}| m{0.72cm}| }
		\hline
		KB No. & $\mathfrak{g}'$ & $\sf Q_{G'}$ & $\mathfrak{g}$ & $\sf Q_{G}$ & Act & Dim \\ \hline	 
	 	 9 & $D_4$ &
		 $  \node{}1 - \node{}2 < \equiv  \nodesq{}3 $  
		 & $G_2$ &
		 $  \node{\flav 1}2 -  \node{}3\supset \footnotesize{adj} $
		 &  $S_3$ &  \footnotesize{$5$}
	 \\ \hline
	\end{tabular}
	\caption{Ninth Kostant-Brylinski Result, Corresponding Coulomb Branch Quivers, Action and Dimension. The $\supset adj$ is used to denote an adjoint node (\cite{Hanany:2018vph,DGHZ18}). The action of the permutation group of three elements is denoted by $S_3$.
	%\textcolor{red}{Should I cite Schlomo and Gaiotto for their G2? Other work?}
	}
	\label{tab:KBninth}
\end{table}
In fact, \ref{eq:ENO} is part of a larger commutative diagram shown in figure \ref{fig:Commuttwistedd}.
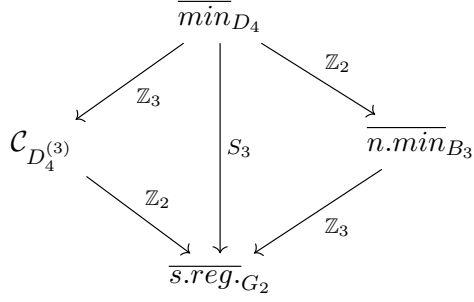
\begin{figure}[h]
\center{
\begin{tikzcd}[row sep=2.5em]
  &  \overline{min}_{D_4}  \arrow{dl}{\mathbb{Z}_3}  \arrow{dd}{S_3} \arrow{dr}{\mathbb{Z}_2} &   \\
  \mathcal{C}_{D_4^{(3)}} \arrow{dr}{\mathbb{Z}_2} & & \overline{n.min}_{B_3} \arrow{dl}{\mathbb{Z}_3}
\\ & \overline{s.reg.}_{G_2} &
\end{tikzcd}
}
\caption{\label{fig:Commuttwistedd} Commutative Diagram of the orbifolding of $\overline{min}_{D_4}$, $\mathcal{C}_{D_4^{(3)}}$, $\overline{n.min}_{B_3}$ and $\overline{s.reg.}_{G_2}$ Coulomb Branches.}
\end{figure}
Except for the newly found $\mathcal{C}_{D_4^{(3)}}$, the commutative diagram is obtained in the analysis of discrete gauging \cite{Hanany:2018vph,DGHZ18}. The quiver for the sub-regular orbit of $G_2$ (depicted on the right side in table \ref{tab:KBninth}) is studied in detail in \cite{Cremonesi:2014vla,Hanany:2018vph,DGHZ18,Hanany:2017ooe} and the HWG has the form:
\be \label{eq:G2subreghwg}
HWG_{\; \overline{s.reg.}_{G_2}}=PE\left[\nu_1 t^2 +{\nu_2}^2 t^4 +{\nu_2}^3 t^6 +{\nu_1}^2 t^8 +{\nu_2}^3 \nu_1 t^{10}-{\nu_2}^6 {\nu_1}^2 t^{20}\right] ,
\ee
where $\nu_1$ and $\nu_2$ denote the highest weight fugacities for the adjoint and fundamental representation of $G_2$, respectively. HWG \ref{eq:G2subreghwg} can be obtained from \ref{eq:tad4hwg} by explicit $\mathbb{Z}_2$ projection which acts non-trivially only on $\nu_2^3 t^6$ and $\nu_1 t^4$ terms:
\be
\begin{split}
HWG_{\; \overline{s.reg.}_{G_2}} = & PE\left[\nu_1 t^2 +\nu_2^2 t^4 + \nu_2^3 t^6 - \nu_2^6 t^{12}\right] \times \\
&  \frac{1}{2} \left(  \frac{1}{\left(1-\nu_1 t^4\right)\left(1-\nu_2^3 t^6\right)} + \frac{1}{\left(1+\nu_1 t^4\right)\left(1+ \nu_2^3 t^6\right)} \right)
\end{split}
\ee
\end{itemize}

%%%%%%% COMPUTATION OF THE B3 to G2 decomposition!!!!!
One could also obtain HWG \ref{eq:G2subreghwg} by starting with Coulomb branch for the closure of the next-to-minimal nilpotent orbit of $B_3$ shown on the right side of the commutative diagram in figure \ref{fig:Commuttwistedd}. The corresponding Highest Weight Generating function takes the form \cite{HK16,DGHZ18}:
%(see e.g. Table 10 in \cite{HK16} or equation 2.77 in \cite{DGHZ18}):
\be \label{eq:b3nminhwg}
HWG_{\: \overline{n.min}_{B_3}}=PE \left[ \lambda_2 t^2 + \lambda_1^2 t^4 \right] ,
\ee
where $\lambda_2$ and $\lambda_1$ are the highest weight fugacities for the adjoint and vector representations of $SO(7)$, respectively.

First, use HWG \ref{eq:b3nminhwg} to obtain the refined Hilbert Series corresponding to $\overline{n.min}_{B_3}$. Furthermore, project from $SO(7)$ to the $G_2$ lattice to obtain a refined HS which is now written in terms of fugacities for the fundamental weights of $G_2$. Using the newly obtained HS, compute the corresponding HWG. In terms of the $G_2$ highest weight fugacities $\nu_1, \nu_2$, one finds
\be \label{eq:g2auxhwg}
HWG_{\; \overline{n.min}_{B_3}} =\frac{1 + \nu_1 \nu_2 t^{6}}{\left(1 - \nu_1 t^2\right) \left(1 - \nu_2 t^2\right) \left(1 - 
   \nu_2^2 t^4\right) \left(1 - \nu_1^2 t^8\right) } .
\ee
Next, to employ the desired $\mathbb{Z}_3$ action using $\omega$, similarly to \ref{eq:z3average} and \ref{eq:d4twistedaverage} preform the averaging of \ref{eq:g2auxhwg}.
\be \label{eq:g2auxhwgaverage}
\frac{1}{3} \sum_{i=0}^{2}  \left( \frac{1 +  \omega^i \nu_1 \nu_2 t^6}{\left(1 - \nu_1 t^2\right) \left(1 -  \omega^i \nu_2 t^2\right) \left(1 -  \nu_2^2 t^4\right) \left(1 - \nu_1^2 t^8\right) } \right)
\ee
As before, the action naturally leaves invariant the terms involving only the adjoint fugacity $\nu_1$. In addition, the $\nu_2^2 t^4$ term is also invariant under the action. Finally, evaluating the $\mathbb{Z}_3$ averaging in \ref{eq:g2auxhwgaverage} yields the HWG given by \ref{eq:G2subreghwg}. \\
%The decomposition of the $B_3$ highest weight fugacity in terms of the $G_2$ fugacities yield 
%\be
%\mu_2 \longrightarrow \mu_1 + \mu_2 .
% PE \left[ \nu_1 t^2 \nu_2 t^2 + \nu_2^2 t^4 + \nu_1 \nu_2 t^6 + \nu_1^2 t^8 -\nu_1^2 \nu_2^2 t^{12} \right]
%\ee
In summary of the last two sections, the Coulomb branch results for the affine $G_2$ and twisted affine $D_4^{(3)}$ quivers invariantly suggest the validity of Claim \ref{Main Claim}. The quivers in table \ref{tab:G2min} and \ref{tab:D4twisted} both lack any outer automorphism symmetry and both have $k_s=1$ (i.e. a single rank $1$ short node). Indeed, as expected, one finds $2$ different Coulomb branches per each quiver. Finally, orbifold relations \ref{eq:KB92resultg2} and \ref{eq:ENO} provide further evidence for Claim \ref{Orbifold Claim}.

%PL of the C_4^(3) is PL= [1,0]t^2 + ([1,0]-[0,0])t^4 - ([1,0]+[0,1]+[2,0]+[0,0])t^6 - (-[2,0]+[1,0]+[0,1])t^8

%%%%%%%% Generalizations and C_n HWG
\section{Ungauging Schemes and HWG for $C_n$ Sequence} \label{7}

After having established Claim \ref{Main Claim} and Claim \ref{Orbifold Claim} in the previous sections, we are in a position to proceed towards some generalizations. Recall that in section \ref{2} two different Coulomb branches have been found for a minimally unbalanced $C_3$ quiver. Let us now consider a more general theory depicted in \ref{eq:Cnquiver} termed the minimally unbalanced $C_n$ sequence. 
\be \label{eq:Cnquiver}
\node{}1- \node{}2-\node{}3 - \dots - \node{}{n-1} <= \node{}n - \noder{}2 
\ee
The various ungauging schemes are collected with their respective Hilbert Series and Coulomb branches in table \ref{tab:CgenTab}. Recall that on the long side of the quiver it suffices to show only one of the long ungauging schemes as the other choices are equivalent. Also note that the invalid ungauging schemes (i.e. ungauging on a rank $r$ short node with $r>1$) are omitted. The Coulomb branches corresponding to the long and short rank $1$ ungauging schemes are denoted by $\mathcal{L}_n$ and $\mathcal{S}_n$, respectively. Moreover, according to Claim \ref{Orbifold Claim} one expects the orbifold relation \ref{eq:cgenorbifold} holds.
\be \label{eq:cgenorbifold}
\mathcal{S}_n =\mathcal{L}_n/ \mathbb{Z}_2.
\ee
 First, lets study the long ungauging scheme for $n=4$.\\

\begin{table}
	\centering
	\begin{tabular}{|c|c|}
		\hline
		Ungauging scheme & Hilbert Series \& Coulomb branch \\ \hline		
$\node{}1- \node{}2 - \dots - \node{}{n-1} <= \node{}n - \nodersq{}2 $
& 
\begin{tabular}{|c|c|c|} \hline
$n$ & HS & $\mathcal{C} $ \\ \hline
3 & $\frac{1}{(1-t)^{14}}$ & $\mathbb{H}^{7}$ \\ \hline
4 & $ \frac{ (1 + 55 t^2+ 890 t^4 + 5886t^6 + 17929t^8 + 26060 t^{10} + palindrome + t^{20} ) }{(1+t^2)^{-1}(1 - t^2)^{22}}$ & $\overline{min}_{E_6}$ \\ \hline
%5 & to be added & $\overline{n.n.n.n.min\;C5}$ \\ \hline
\end{tabular}
 \\ \hline \hline
$\nodesq{}1- \node{}2 - \dots - \node{}{n-1} <= \node{}n - \noder{}2 $
& 
\begin{tabular}{|c|c|c|} \hline
$n$ & HS & $\mathcal{C} $ \\ \hline
3 & $\frac{1+6t^2 +t^4}{(1 - t)^{10}(1-t^2)^4}$ & $\mathbb{H}^{5}\times \mathbb{C}^4/\mathbb{Z}_2$ \\ \hline
4 & $ \frac{ (1 + 29 t^2+ 435 t^4 + 2948t^6 + 8998t^8 + 12969 t^{10} + palindrome + t^{20} ) }{(1+t^2)^{-1} (1 - t^2)^{22}}$ & $\overline{n.min }_{F_4}$ \\ \hline
%5 & to be added & $\overline{C_5} / \mathbb{Z}_2$ \\ \hline
\end{tabular}
 \\ \hline
%$\node{}1- \nodesq{}2-\node{}3 - \dots - \node{}{n-1} <= \node{}n - \noder{}2 $ &
%%& $HS(t)=\frac{(1 + t^2) (1 + 6 t + 18 t^2 + 28 t^3 + 38 t^4 + 28 t^5 + 18 t^6 + 
%%   6 t^7 + t^8)}{(-1 + t)^{n^2-3} (1 - t^2)^6 (1 + t + t^2)^2}, \; 
%$\mathcal{C}=$ 
%not valid 
%\\ \hline
%
%$\node{}1- \node{}2-\node{}3 - \dots - \nodesq{}{n-1} <= \node{}n - \noder{}2 $
%& $ \mathcal{C}=$not valid \\ \hline
	\end{tabular}
	\caption{ Different choices of ungauging schemes for the minimally unbalanced $C_n$ sequence.}
	\label{tab:CgenTab}
\end{table}
Observe that for $n=4$ one obtains the twisted affine $E_6^{(2)}$ quiver with Coulomb branch corresponding to the space of one $E_6$ instanton on $\mathbb{C}^2$, or equivalently, to the closure of the minimal nilpotent orbit of $\mathfrak{e}_6$ algebra \cite{Hanany:2017ooe}. The Highest Weight Generating function written in terms of the highest weight fugacity for the adjoint representation of $E_6$ takes the simple form \cite{Hanany:2017ooe}: \\
\be \label{eq:e6hwg}
HWG_{\; \overline{min}_{E_6}} = PE\left[\lambda_6 t^2\right] .
\ee
The ungauging scheme in the lower part of table \ref{tab:CgenTab} produces an orbifold Coulomb branch (see section \ref{2} for the case $n=3$). For $n=4$, one obtains the closure of the next-to-minimal nilpotent orbit of $\mathfrak{f_4}$ algebra, with the Highest Weight Generating function \cite{Hanany:2017ooe}:
\be \label{eq:f4nmin}
HWG_{\; \overline{n.min}_{F_4}} = PE\left[\nu_1 t^2 + \nu_4^2 t^4\right] ,
\ee
where $\nu_1$ and  $\nu_4$ are the highest weight fugacities for the adjoint and fundamental representations of $F_4$, respectively. To identify the $\mathbb{Z}_2$ action explicitly, first decompose the highest weight fugacity for the adjoint of $E_6$ in terms of the $F_4$ fugacities:
\be
\lambda_6 \longrightarrow \nu_1 + \nu_4 .
\ee
The $\mathbb{Z}_2$ projection is then constructed as
\be
\frac{1}{2}\left( \frac{1}{\left(1-\nu_1 t^2\right) \left(1- \nu_4 t^2\right)} + \frac{1}{
\left( 1-\nu_1 t^2\right) \left(1+ \nu_4 t^2\right)} \right)
\ee
which indeed equals the Highest Weight Generating function \ref{eq:f4nmin}. Hence, in accord with Claim \ref{Orbifold Claim}, one finds the orbifold relation \ref{eq:e6f4}.
\be \label{eq:e6f4}
 \overline{n.min}_{F_4}=\overline{min}_{E_6}/ \mathbb{Z}_2
\ee
Equation \ref{eq:e6f4} reproduces the fourth result of Kostant and Brylinski given in the first row of table \ref{tab:KB} in the introduction. Further treatment of this case should appear in \cite{BHMfolding20}. \\

Now, lets turn our attention back to the long unaguging scheme in table \ref{tab:CgenTab} and lets derive the HWG in terms of the highest weight fugacities of $C_4$. Begin with HWG \ref{eq:e6hwg} and compute the decompositions of the $E_6$ adjoint highest weight fugacity as well as its second and third power:
\be
\begin{split}
\lambda_6 & \longrightarrow \mu_1^2 + \mu_4 ,\\
 \lambda_6^2 & \longrightarrow \mu_1^4 + \mu_1^2 \mu_4 + \mu_4^2 + \mu_2^2 + 1 + \mu_4, \\
 \lambda_6^3 & \longrightarrow \mu_1^6 +\mu_1^4 \mu_4 + \mu_1^2 \mu_4^2 +\mu_4^3 +\left(\mu_2^2 +1+\mu_4 \right)\left(\mu_1^2 +\mu_4\right) + \mu_3^2 ,
 \end{split}
 \ee
where $\mu_1, \mu_2, \mu_3, \mu_4$ denote the highest weight fugacities of $C_4$. As a result, one obtains the Highest Weight Generating function in terms of the $C_4$ highest weight fugacities:
\be \label{eq:c4hwg}
HWG_{\; \mathcal{L}_4}=PE\left[ \mu_1^2 t^2  + \mu_2^2 t^4 + \mu_3^2 t^6 + t^4 + \mu_4 t^2  + \mu_4 t^4 \right] .
\ee
Inspection of HWG \ref{eq:hwgC3} in section \ref{2} and HWG \ref{eq:c4hwg} suggests generalization for any value of $n$ which is given by \ref{eq:hwgCgen}.
\be \label{eq:hwgCgen}
HWG_{\; \mathcal{L}_n}=PE\left[\sum_{i=1}^{n-1} \mu_i^2 t^{2i} + t^4 +\mu_n\left(t^{n-2}+t^n\right)\right],
\ee 
where $[\mu_1,\dots, \mu_n]$ denote the fugacities for the highest weights of $Sp(n)$. Prediction \ref{eq:hwgCgen} has been tested for $n$ up to $5$. Note that HWG \ref{eq:hwgCgen} can also be obtained as a `folding' of the quiver in Figure 4 of \cite{HPsic17} upon setting $N=4$. A general formula for the HWG for the quotient space $\mathcal{S}_n$, corresponding to the short ungauging scheme depicted in the lower part of table \ref{tab:CgenTab}, turns out considerably more difficult compared to \ref{eq:hwgCgen} due to smaller global symmetry of the Coulomb branch. 
%\begin{table}
%	\centering
%\begin{tabular}{| m{0.43cm} |m{0.65cm}| m{3cm} | m{0.43cm} | m{3cm}| m{0.5cm}| m{0.72cm}| }
%		\hline
%		KB No. & $\mathfrak{g}'$ & $\sf Q_{G'}$ & $\mathfrak{g}$ & $\sf Q_{G}$ & Act & Dim \\ \hline	 
%	 	 4 & $E_6$ &
%		 $ \node{}1 - \node{}2 - \node{\topnode{\topnode{}1}2}3 - \node{}2 - \node{}1 $ 
%		 & $F_4$ &
%		  $ \node{}1 - \node{}2 - \node{}3 <= \node{}4 - \node{}2 $
%		   & \footnotesize{DU}   &  \footnotesize{$11$}
%	 \\ \hline	
%	\end{tabular}
%	\caption{Fourth Kostant-Brylinski Result, Corresponding Coulomb Branch Quivers, Action and Dimension. DU denotes an action on quivers termed \emph{discrete ungauging} which is studied in \cite{BHMfolding20}.
%	}
%	\label{tab:KBe6f4}
%\end{table}

%Do unrefined from general HWG and subtract. so go to higher values of n and higher orders in the perturb expansion.

% Do \mathcal{S}_n = Have to re-express in terms of C_{n-1}.
% n=3 decomposition into C2
% HWG_{\mathcal{L}_3}= PE \left[ \mu_1^2 t^2 + \mu_2^2 t^4 + t^4+ \mu_3 t + \mu_3 t^3 \right] (in C3)
% HWG_{\mathcal{S}_3}= PE \left[ \mu_1 t + \mu_1^2 t^2 \right] (in C2)

% HWG_{\mathcal{L}_4}= PE\left[ \mu_1^2 t^2  + \mu_2^2 t^4 + \mu_3^2 t^6 + t^4 + \mu_4 t^2  + \mu_4 t^4 \right]  (in C4)
%HWG_{\mathcal{S}_4}=PE\left[\nu_1 t^2 + \nu_4^2 t^4\right] (in F4)
%decompose F4 into C3:

%\nu_1 \longrightarrow 

% HWG_{\mathcal{S}_4}= PE \left[ ] (in C3)?

%%%%%%%%%%%%%%%%%%%%%           Conclusions & Prospects    
%%%%%%%%%%%%%%%%%%%%%%%%%%%
\section{Conclusions and Prospects} \label{conclusions}

This note demonstrates that the Coulomb branch of a non-simply laced quiver theory varies depending on the ungauging scheme (i.e. the choice of node where a $U(1)$ symmetry is ungauged). All ungauging schemes on the long side of the quiver yield the same Coulomb branch $\mathcal{C}_L$ described by the same Hilbert Series. Short ungauging schemes (i.e. those which involve ungauging on short nodes of the quiver) do not in general correspond to a valid Coulomb branch (See Appendix \ref{B}) except when they involve ungauging on a rank $1$ node, in which case, the Coulomb branch takes the form:
\be
\mathcal{C}_S= \mathcal{C}_L / \mathbb{Z}_k ,
\ee
where $k$ is the multiplicity of the non-simply laced edge (i.e. $n=2$ for double-laced edge, $n=3$ for triple-laced edge, and so on). Generally, for quivers in the form of an affine Dynkin digram, ungauging the affine node (which is long in all non-simply laced cases) leads to the simplest Coulomb branch - corresponding to both the reduced moduli space of one $G$ instanton on $\mathbb{C}^2$ as well as to the closure of the minimal nilpotent orbit of the corresponding Lie algebra $\mathfrak{g}$. \\

Remarkably, the ungauging scheme analysis reproduces the mathematical results of Kostant-Brylinski \cite{KB92} contained in table \ref{tab:KB} in terms of $3$d Coulomb branch quivers. Generally, Coulomb branch quivers provide intuitive ways to understand the orbifold relations between moduli spaces thanks to their graph theoretic nature and owing to the powerful methods developed for the study of moduli spaces of supersymmetric gauge theories.\\

Follow-up future work might include:
\begin{itemize}
\item Understanding the remaining results of Kostant-Brylinski \cite{KB92} in terms of three-dimensional Coulomb branch quivers \cite{BHMfolding20}.
\item Carrying out the ungauging scheme analysis for all minimally unbalanced quivers \cite{MUQ18} with single non-simply laced edge and at least one rank $1$ short node.  Furthermore, one can expand the ungauging scheme analysis to quivers with two non-simply laced edges. For the former and latter purposes, one could start amid quivers contained in the classification of simply-laced and the exotic classification of non-simply laced minimally unbalanced quivers \cite{MUQ18}, respectively. Based on a few studied cases, we suspect that the orbifold relations are transitive. In particular, the schematic example in table \ref{tab:speculation} illustrates the transitive behavior of the two ungauging schemes for a quiver with two non-simply laced edges.
\begin{table} \label{tab:KBninth}
	\centering
\begin{tabular}{| m{4cm} |m{3cm}|}
		\hline
		Ungauging Scheme & Coulomb Branch \\ \hline	 
	$\nodesq{}1=>\node{}2 - \node{}3 => \node{}2 - \node{}1$ & $\mathcal{C}$ \\ \hline
	$\node{}1=>\node{}2 - \node{}3 => \node{}2 - \nodesq{}1$ & $\mathcal{C}/ \left( \mathbb{Z}_2 \times \mathbb{Z}_2 \right)$
	 \\ \hline
	\end{tabular}
	\caption{Transitive orbifold behavior of Coulomb branches for quiver with two non-simply laced edges.
	%\textcolor{red}{Should I cite Schlomo and Gaiotto for their G2? Other work?}
	}
	\label{tab:speculation}
\end{table}
%By doing so, one can exhaust all cases of Coulomb branches with finite Dynkin diagram global symmetry which are related by an orbifold action. Such classification is the subject of ongoing development \cite{HZorbifoldclassification20}.
% \item Carrying out the ungauging scheme analysis for quivers with two or more non-simply laced edges. For this purpose, one could start amid quivers contained in the exotic classification of non-simply laced minimally unbalanced quivers \cite{MUQ18}. 
%\item Analysis of this phenomenon for quivers with adjoint loops which naturally appear in the $3d$ description of the $6d$ world-volume theories of stacks of five-branes on orbifold singularities in M-theory.
%\item Detailed analysis of what happens to magnetic lattices 
\end{itemize}
As hinted by result \ref{eq:ENO}, there are more cases of orbifold relations between three-dimensional Coulomb branches yet to be discovered.

\section*{Acknowledgements}
This work is supported by STFC Consolidated Grant ST/P000762/1. A.H. and A.Z. would like to thank the organizers of the 2019 Pollica Summer Workshop on Mathematical and Geometric Tools for Conformal Field Theories for their hospitality. A.Z. would like to extend his gratitude to Sun Woo Kim for discussions which sprang the initiation of this project. A.Z. would like to thank Rudolph Kalveks for especially helpful discussions and succour during the course of this project. A.Z. is also thankful to Antoine Bourget and Dominik Miketa for enlightening discussions. 

\appendix

\section{Monopole Formula} \label{A}
Moduli spaces of $3d \ \mathcal N=4$ quiver gauge theories have two distinct phases known as the \emph{Coulomb branch} (where the gauge group is broken to its maximal torus) and \emph{Higgs branch} (where the gauge group of the theory is either fully or partially broken). When the scalars in the vector multiplet assume vacuum expectation values (vevs) and the scalars in the hypermultiplet are all zero, we are on the Coulomb branch of the moduli space. On the other hand, if only the scalars from hypermultiplets assume non-zero vevs and the vector multiplet scalar vevs are vanishing one probes the Higgs branch of the moduli space.  For most good theories in the sense of \cite{GW09}  the \emph{Coulomb branch} (and also the \emph{Higgs branch}) is a hyperK\"ahler  variety which can be described by its ring of holomorphic functions. The information about the variety is encoded in a Hilbert Series which succinctly enumerates holomorphic functions in the coordinate ring. A one-to-one correspondence exists between holomorphic functions in the moduli space and gauge invariant BPS operators in the chiral ring of a quantum field theory. The \emph{monopole formula} used for counting these operators, developed in \cite{CHZ13}, takes the form:
\begin{equation} \label{eq:monopoleformula}
H_G(t,z)=\sum_{m \in \Gamma _{\hat{G}} / \mathcal{W}_{\hat{G}}} z^{J(m)} t^{\Delta(m)} P_{G} (t,m) ,
\end{equation}
where $G$ is the gauge group of the theory and $m$ denotes the magnetic charges (see \cite{GNO76}) which take values in the magnetic lattice:
\begin{align} \label{eq:lattice}
\Gamma_{\hat {G}}:=(\Gamma_{G})^* ,
\end{align}
where $(\Gamma_{G})^*$ is the lattice dual to the weight lattice of $G$. It defines a new weight lattice of a new group $\hat G$, which is the GNO dual of $G$ \cite{GNO76}. $\mathcal{W}_{\hat{G}}$ is the Weyl group of $\hat G$. $J(m)$ denotes the topological charge counted by the fugacity $z$. The dressing factor $P_G$ is a generating function for Casimir invariants of the unbroken gauge group. $\Delta(m)$ is the conformal dimension which coincides with the R-charge of the monopole operators. The conformal dimension, obtained using a radial quantization is \cite{Borokhov:2002ib}:
\begin{equation} \label{eq:cd}
\Delta(m)= \Delta(m)_V + \Delta(m)_H =  -\sum_{\alpha \in \Delta_+} \mid\alpha(m)\mid +\frac{1}{2} \sum_{i=1} ^n \sum_{\rho_i \in R_i} \mid\rho_i (m)\mid .
\end{equation}
The two terms of the conformal dimension formula $\Delta(m)_V$ and  $\Delta(m)_H$ account for vector multiplet and hypermultiplet contributions, respectively. $\Delta_+$ is the set of positive roots of the gauge group. Hypermultiplets transform in representations $R_i$ with weights $\rho_i$. For a more detailed exposition of the monopole formula, see \cite{CHZ13}. To treat non-simply laced quivers, a modification of the hypermultiplet contribution to the conformal dimension $\Delta(m)_H$, introduced in \cite{CFHM14}, takes the following form:
\begin{equation}
\frac{1}{2}  \mid\rho_i (m)\mid \rightarrow \frac{1}{2} \sum_{j=1} ^{N_1} \sum_{k=1} ^{N_{2}} \mid \lambda m_j ^{(1)} - m_k ^{(2)}\mid ,
\end{equation}
where $\rho_i$ is the irrep corresponding to the hypermultiplets assigned to the edge between two nodes $U(N_1)$ and $U(N_2)$. Setting $\lambda =1$ recovers the formula for a simple laced quiver, $\lambda =2$ is used for a double laced edge, $\lambda =3$ for a triple laced edge, and so on. The direction of the edge points from $N_1$ to $N_2$. $m^{(1)}$ and $m^{(2)}$ denote the magnetic fluxes for $U(N_1)$ and $U(N_2)$, respectively. The function which enumerates the Casimir invariants of the residual gauge group of $U(N)$, which is left unbroken by the configuration of magnetic charges, has the form:
\begin{equation}
P_{U(N)} (t;m) =\prod_{k=1} ^N \frac{1}{(1-t^{2k})^{{\sigma}(k)(m)}} ,
\end{equation}
where ${\sigma}(k)({m})$ encodes the various configurations of the gauge symmetry breaking in form of a partition. As a simple example, for a $U(2)$ gauge symmetry and magnetic charges ${m} = (m_1,m_2)$ the dressing factor is:
\begin{equation}
P_{U(2)} (t;m_1,m_2) =
\begin{cases}
               \frac{1}{(1-t)(1-t^{2})} \quad if\quad m_1=m_2 \\
          \frac{1}{(1-t)(1-t)}\quad  if\quad m_1 \neq m_2 .
            \end{cases} 
\end{equation}
In order to use the monopole formula, there are certain restrictions for the conformal dimension, yielding the good, bad, or ugly classification of $3$d $\mathcal{N}=4$ gauge theories of Gaiotto-Witten \cite{GW09}. These restrictions translate into the balancing of the quiver nodes. In particular, for ADE quivers, the balance of a particular $U(N_i)$ gauge node is defined as \cite{HK16}:

\begin{equation} \label{eq:balance}
Balance_{ADE}(i) = \sum_{j\in \: adjecent \: nodes} N_j - 2N_i .
\end{equation}
For BCF and G quivers, the long node directly adjacent to the NSL edge gets double and triple the contribution from the node on the other side of the NSL edge, respectively. A quiver is said to be balanced iff the balance of all its nodes is zero. If one or more nodes in the quiver have positive balance the quiver is said to be positively balanced. A quiver with a single unbalanced node is termed \emph{minimally unbalanced} \cite{MUQ18}. The present work only concerns balanced and minimally unbalanced quivers. For such quivers the conformal dimension satisfies $\Delta(m)>0$ for all $m\in \Gamma _{\hat G}$, which guarantees that the monopole formula can be applied to calculate the Coulomb branch of the moduli space.\footnote{In fact, there are examples of balanced quivers with moduli spaces that are not hyperK\"ahler cones, hence the monopole formula falls short. Hence, it seems that balance is necessary but not sufficient condition for a quiver to be well behaved and treatable by monopole formula methods. Prominent examples of such quivers are multiples of affine Dynkin diagrams.} \\

\section{Choice of Ungauging Scheme and the Conformal Dimension} \label{B}
Lets show how choosing a particular ungauging scheme effects the monopole formula calculation of the Coulomb branch. The monopole formula \ref{eq:monopoleformula} contains a sum over magnetic charges which take values in the lattice \ref{eq:lattice} (see Appendix \ref{A}). The difference between two ungauging schemes corresponds to a shift in the values of magnetic charges or, in other words, to the change of the magnetic lattice over which the summons run in \ref{eq:monopoleformula}. The dressing factors $P_{G} (t,m)$ in \ref{eq:monopoleformula}  are invariant under any shifts in $m$. Conformal dimension $\Delta(m)$, spelled out in \ref{eq:cd}, is the only part that is affected. Furthermore, the vector multiplet contribution $\Delta(m)_V$ is invariant under shifts of $m$ and only $\Delta(m)_H$ changes nontrivialy. To see the effect, consider the quiver \ref{eq:q1}, where $k$ denotes the multiplicity of the non-simply laced edge and $a,b,c,d,e$ denote the magnetic flux vectors at the corresponding nodes. Let the ranks of the nodes be $r_i$, where $i=a,b,c,d,e$.  %how about topological charges?
\begin{equation} \label{eq:q1}
\node{}a-\node{}b-\node{}c < \overset{k}{=}\node{}d-\nodesq{}e
\end{equation}
The relevant contribution to the conformal dimension formula for quiver \ref{eq:q1} has the form
\be \label{eq:deltaC}
\Delta(a,b,c,d,e)_H= \sum \left( |a-b| +| b-c| +|c-kd| +|d-e| \right) \times \delta(e_1) ,
\ee
where $\delta(e_1)$ signifies that the ungauging scheme requires one of the magnetic charges on the long $e$ node to be set to zero. To see what happens when a different ungauging scheme is chosen such that we ungauge on the $d$ node, lets   shift the magnetic charge vectors
\be
e \rightarrow e+ d_1
\ee
resulting in the following form of the terms in the conformal dimension
\be
\left(  |a-b | + |b-c| +|c-kd |+|d-e-d_1|\right)  \times \delta(e_1+d_1).
\ee
Now, shifting $d \rightarrow d- e_1$ and leaving out $d_1$ which is guaranteed to be zero because of the delta function one arrives at
\be \label{eq:deltad}
\Delta_H=\sum \left( |a-b| + |b-c| +|c-kd| +|d-e| \right) \times \delta(d_1),
\ee
which is readily identified as the conformal dimension corresponding to the ungauging scheme for which one ungauges on the long $d$ node. This shows that the Hilbert series and hence the Coulomb branch is the same for both ungauging schemes. Carrying this analysis for a generic quiver provides a proof that all ungauging schlemes on the long side of a non-simply laced quiver yield the same Coulomb branch denoted by $\mathcal{C}_L$. This proves fully Claim \ref{Claim 2} and partially Claim \ref{Main Claim}.
To see the effect of 'jumping' with the ungauging schemes over the non-simply laced edge, perform a shift $d \rightarrow d+ c_1$. One obtains
\be
\left( |a-b| + |b-c| +|c-k(d+c_1)| +|d-e + c_1 |\right) \times \delta(d_1+c_1).
\ee
Next, make the shifts
 \be
  a \rightarrow a- d_1,
   b \rightarrow b-d_1,
 c \rightarrow c- d_1, 
  e \rightarrow e- d_1
\ee
 to get
\be \label{eq:deltac}
\left( |a-b| + |b-c| +|c-kd+(k-1)d_1| +|d-e| \right) \times \delta(c_1).
\ee
Note, that for a simply laced quiver, $k=1$, and expression \ref{eq:deltac} reduces to expression \ref{eq:deltaC} with $c_1 \longleftrightarrow e_1$. Hence, ungauging on the $c$ node would yield the same Coulomb branch. Lets proceed to see how the ungauging on the $b$ node changes the expression for conformal dimension. Performing the shifts
\be
a \rightarrow a - c_1, \;
b \rightarrow b- c_1, \;
c\rightarrow c+b_1-c_1, \;
d \rightarrow d - c_1, \;
e\rightarrow e - c_1,
\ee
and taking into account that the delta function sets $b_1$ to zero, results in the hypermultiplet contribution of the conformal dimension associated to the ungauging on the $b$ node
\be \label{eq:deltab}
\Delta_H=\sum \left( |a-b| + |b-c| +|c-kd+(k-1)d_1| +|d-e| \right) \times \delta(b_1).
\ee
Finally, shift
\be
a \rightarrow a- b_1, \;
b \rightarrow b+a_1 -b_1, \;
c \rightarrow c-b_1, \;
d \rightarrow d-b_1, \;
e \rightarrow e-b_1,
\ee
taking into account that $\delta(a_1)$ sets $a_1$ zero to obtain
\be \label{eq:deltaa}
\Delta_H=\sum \left( |a-b| + |b-c| +|c-kd+(k-1)d_1| +|d-e|\right) \times \delta(a_1).
\ee
Expression \ref{eq:deltaa} shows the structure of the hypermultiplet contribution in the conformal dimension for ungauging scheme on the leftmost $a$ node of the quiver \ref{eq:q1}.\\

The difference between the hypermultiplet contributions to the conformal dimensions for the ungauging schemes on the long side versus on the short side of the quiver i.e. \ref{eq:deltaC}, \ref{eq:deltad} versus \ref{eq:deltac}, \ref{eq:deltab} and \ref{eq:deltaa} boils down to the scaling of one of the axis of the discrete magnetic lattice at the respective node. In general, the magnetic lattice over which the monopole formula summation runs is squashed by this scaling in a non-conformal manner and produces a lattice that does not belong to any family of semi-simple  Lie algebra lattices - hence it is not a valid magnetic lattice of the gauge group at the particular node. This means that the Coulomb branch for such choices of ungauging schemes is not a well defined object. \\
However, if a short rank $1$ node is ungauged (i.e. $r_a=1$), the discrete magnetic lattice at that node is one-dimensional to start with, in particular, it is $\mathbb{Z}$ since the gauge group is $U(1)$. The effect of the rescaling of the lattice due to the choice of short ungauging scheme is that: $\mathbb{Z} \longrightarrow k\mathbb{Z}$, where $k$ is the multiplicity of the non-simply laced edge. Now, the monopole formula summation runs over every $k$-th point compared to the summation for any long ungauging scheme. Hence, the computed Coulomb branch is of the form:
\be
\mathcal{C}_S = \mathcal{C}_L / \mathbb{Z}_k .
\ee
This provides a schematic proof of Claim \ref{Orbifold Claim}.

\bibliography{mainC}
\bibliographystyle{JHEP}

\end{document}